\documentclass{iaus}
\usepackage{graphicx}

\title[X-ray radiation of SS~433] 
{X-ray radiation of the jets and the supercritical accretion disk in SS~433}
\author[S. Fabrika \& A. Medvedev]   
{Sergei Fabrika$^1$
 \and Alexei Medvedev$^2$}

\affiliation{$^1$Special Astrophysical Observatory, \\ 369167, Nizhij Arkhyz, Russia\\
email: {\tt fabrika@sao.ru} \\[\affilskip]
$^2$Moscow State University, \\ 119992, Moscow, Russia \\
email: {\tt a.s.medvedev@gmail.com}}

\pubyear{2010}
\volume{275}  
\pagerange{xxx--xxx}
\jname{Jets at all Scales}
\editors{G.E. Romero, R.A. Sunyaev \& T. Belloni, eds.}
\begin{document}

\maketitle

\begin{abstract} The observed X-ray luminosity of SS~433 is $\sim 10^{36}$\,erg/s, it is known that
all the radiation is formed in the famous SS~433 jets. The bolometric luminosity of SS~433
is $\sim 10^{40}$\,erg/s, and originally the luminosity must be realized in X-rays. The original radiation is 
probably thermalized in the supercritical accretion disk wind, however the missing more than four orders of magnitude
is surprising. We have analysed the XMM-Newton spectra of SS\,433 using a model of
adiabatically and radiatively cooling X-ray jets. The multi-temperature thermal
jet model reproduces very well the strongest observed emission lines, but it can not reproduce the continuum radiation
and some spectral features. We have found a notable contribution of ionized reflection to the spectrum in the energy range
from $\sim 3$ to 12~keV. The reflected spectrum is an evidence of the supercritical disk funnel, where the illuminating
radiation comes from deeper funnel regions, to be further reflected in the outer
visible funnel walls ($r\ge 2 \cdot 10^{11}$ cm). The illuminating spectrum is similar to that observed in ULXs, its 
luminosity has to be no less than  $\sim 10^{39}$\,erg/s. A soft excess has been  detected, that does not depend
on the thermal jet model details. It may be represented as a BB with a temperature of T$_{bb} \approx 0.1$~keV
and luminosity of L$_{bb} \sim 3 \cdot 10^{37}$\,erg/s. The soft spectral component has about the same parameters
as those found in ULXs.
\end{abstract}

SS\,433 is the only known persistent superaccretor in the Galaxy -- a source of 
relativistic jets (\cite[Fabrika (2004)]{Fab04} for review). 
This is a massive close binary, where the compact star is most probably a black hole.
Its intrinsic luminosity is estimated to be $\sim 10^{40}$\,erg/s, with
its maximum located in non-observed UV region. Almost all the observed radiation
is formed in the supercritical accretion disk, and the donor star contributes 
less than 20\,\% of the optical radiation. 
The extreme luminosity of the object is supported by a very well 
measured kinetic luminosity of the jets, $\sim 10^{39}$\,erg/s, both in
direct X-ray and optical studies of the jets and
in the studies of the jet-powered nebula W\,50. 
At the same time we know that practically all the energy at the accretion onto 
a relativistic star is released in X-rays. This means that the observed 
radiation of SS\,433 is a result of thermalization of the original radiation 
in the strong wind forming in the supercritical disk.

The observed X-ray luminosity of SS\,433 is $\sim 10^{36}$\,erg/s and
it is believed that all the X-rays come from the cooling X-ray jets (\cite[Kotani \etal\ 1996]{kotani96}; 
\cite[Marshall, Canizares, \& Schulz 2002]{marshall02}; \cite[Brinkmann, Kotani, \& Kawai 2005]{brinkmann05}). 
The observed X-ray radiation is at least four orders of magnitude less than the 
bolometric luminosity, 
which originally must be released in X-rays. Both the orientation of SS\,433 and the 
visibility conditions do not allow us to see any deep areas of the supercritical accretion disk funnel. However,
the strong mass loss of the accretion disk gives us a hope for detecting some 
indications of the funnel radiation. The goal of our study is to find these indications of the 
funnel in the SS\,433 X-ray radiation.

\begin{figure}
\begin{center}
 \includegraphics[angle=270,width=2.63in]{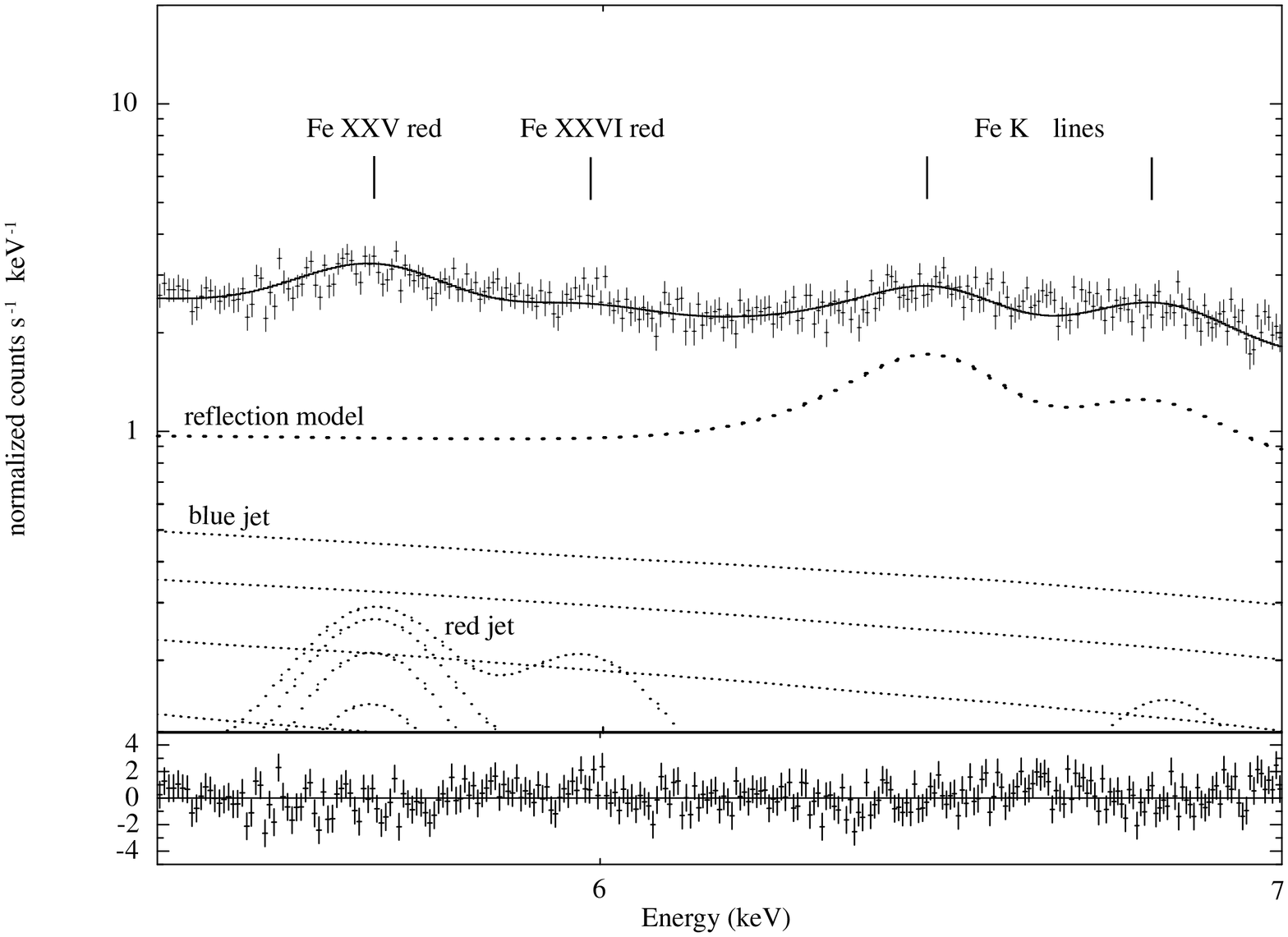} 
\includegraphics[angle=270,width=2.63in]{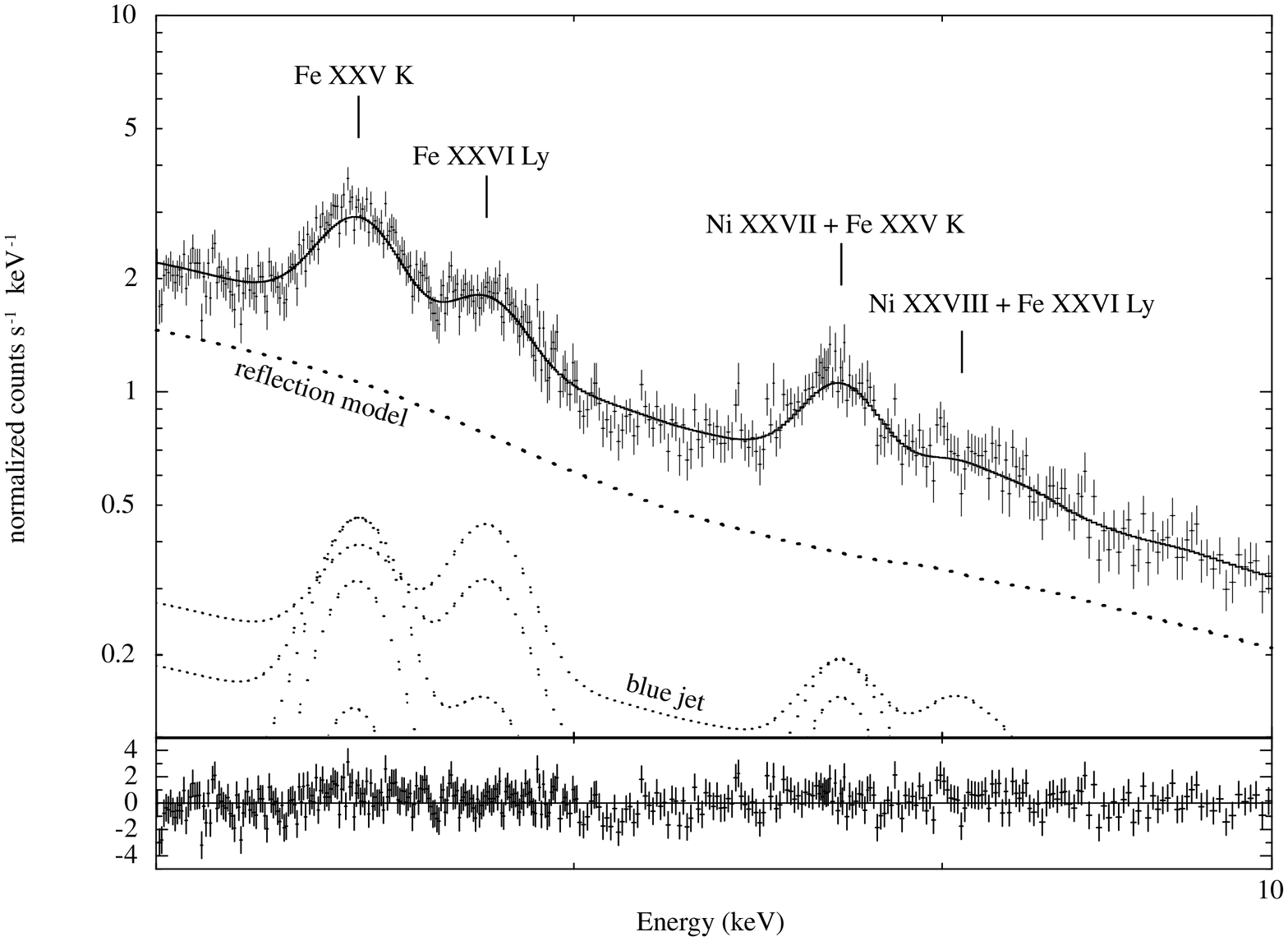}
 \caption{Observed spectrum of SS\,433 (orbit 707, $\psi \sim 0$) 
with the thermal jet model components and the additional reflection model 
in two spectral regions 5.0--7.0 (left) and 7.0--10.0~keV (right). The total
model spectrum is shown with solid line. The reflection model 
explains well the fluorescence iron line at 6.4~keV, the recombination iron 
K$\alpha$ line of Fe\,XXV at 6.7~keV and the iron absorption 
edge (\cite[Kubota \etal\ 2007]{kubota07}) at $\sim$8--9~keV.}
   \label{fig1}
\end{center}
\end{figure}

\begin{figure}[b]
\vspace*{1.0 cm}
\begin{center}
 \includegraphics[width=1.8in]{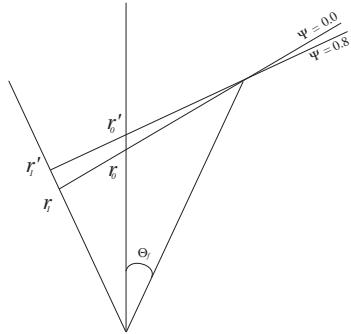} 
 \caption{A sketch of the funnel and the blue jet visibility at two precessional phases
 considered. $r_0$ ($r_0^{\prime}$) is the distance between the base of the visible jet 
 and the top of the cone (the bottom of the funnel). $r_1$ ($r_1^{\prime}$) is the distance 
 between the base of the visible funnel wall and the top of the cone.}
   \label{fig2}
\end{center}
\end{figure}

\begin{figure}
\vspace*{1.0 cm}
\begin{center}
 \includegraphics[width=3.4in]{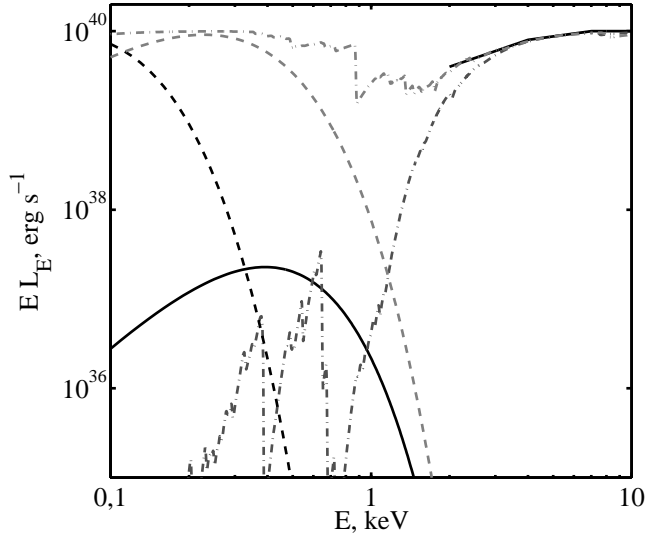} 
 \caption{Additional components in X-ray spectrum of SS\,433 (the thermal jet 
component is not shown). The illuminating incident radiation found in reflection
model is shown by solid lines (arbitrary scaled to $E L_E = 10^{40}$\,erg/s) in three
energy ranges: 2.0\,--\,4.0~keV ($\Gamma = 1.0$), 4.0\,--\,7.0~keV ($\Gamma = 1.6$)
and 7.0\,--\,12.0~keV ($\Gamma = 2.0$). The soft BB component 
($T_{bb} = 0.1$~keV, $L_x \sim 3 \cdot 10^{37}$\,erg/s) 
detected in the 0.8\,--\,2.0~keV energy range is shown 
by solid line. There are two originally the same flat spectra after absorption 
in ABSORI model with $N_H =10.0\times 10^{22}$\,cm$^{-2}$,
$\xi = 150$ (grey dash-dotted line) and  $N_H =6.0 \times 10^{22}$\,cm$^{-2}$,
$\xi = 3$ (bold dash-dotted line). There are two MCF (multi-colour funnel model) 
non-absorbed spectra with $r_1 = 1.7 \cdot 10^{11}$\,cm (bold dotted 
line) and $r_1 = 5 \cdot 10^{10}$\,cm (grey dotted line).}
   \label{fig3}
\end{center}
\end{figure}

We have analysed XXM spectra of SS\,433 with a well-known standard model of the adiabatically 
cooling X-ray jets, taking into account cooling by radiation (\cite[Medvedev \& Fabrika 2010]{Medv2010}).
We have selected only those observations (\cite[Brinkmann, Kotani, \& Kawai 2005]{brinkmann05}) with highest S/N, where 
disk was not eclipsed by the donor and was the most open to the observer (precessional phases $\psi = 0.04$ and 0.84).
We confirm that the jet model reproduces the iron emission line fluxes quite well.
However, the thermal jet model alone can not reproduce the continuum radiation
in the XMM spectral range. We use then the multi-temperature thermal jet model together with 
the REFLION ionized reflection model (\cite[Ross \& Fabian 2005]{ross05}). 
We divide the whole spectrum into four parts (0.8\,--\,2.0, 2.0\,--\,4.0, 4.0\,--\,7.0 
and 7.0\,--\,12.0~keV) and find the reflection model parameters independently in each part,
but with the same jet parameters and the same IS absorption $N_H$. 

Fig.\,\ref{fig1} shows two spectral regions with the "thermal jet + reflection" model, we 
find quite a good representation of the spectra. The reflection model explains well the spectral features
unexplained before. We find that the ionization parameter 
is about the same in the different X-ray ranges, $\xi \sim 300$, which indicates a highly
ionized reflection surface. The illuminating radiation photon index changes 
from flat, $\Gamma \approx 2$, in the 7\,--\,12~keV range to $\Gamma \approx 1.6$
in the range 4\,--\,7~keV and to $\Gamma \le 1$ in the range 2\,--\,4~keV.

Fig.\,\ref{fig2} shows the jet and the probable funnel wall visibility at the 
two precessional phases studied. We found the visible blue jet base is
$r_{0} \approx 2\cdot10^{11}$\,cm, the gas temperature at $r_{0}$ is 
$T_0 \approx 17$~keV. The IS gas column density is 
$N_H \sim 1.5 \cdot 10^{22}$\,cm$^{-2}$, which is in good agreement with the IS
extinction found in optical and UV observations. We confirm the previous finding 
(\cite[Kotani \etal\ 1996]{kotani96}; \cite[Brinkmann, Kotani, \& Kawai 2005]{brinkmann05}) 
that the red jet is probably not seen (blocked)
in the soft energy range of 0.8\,--\,2.0~keV. However, both the model with blocked
the red jet portions, and the model with the whole red jet visible (but with some 
higher $N_H$) give the same result, that the thermal jet model alone can not 
explain the soft continuum. We also confirm that Nickel is highly 
overabundant, at least 10 times, in the jets (\cite[Kotani \etal\ 1996]{kotani96}; 
\cite[Brinkmann, Kotani, \& Kawai 2005]{brinkmann05}). 

We have not found any evidences of the reflection in the soft 0.8\,--\,2.0~keV
energy range, instead the soft excess is detected in spectra. The soft excess
does not depend on the thermal jet model details, this model alone can not 
explain the soft continuum. We suppose that in the soft X-rays we observe direct 
radiation of the visible funnel wall. We represented this component 
(Fig.\,\ref{fig3}) as a black body radiation with a temperature of
$T_{bb} \approx 0.1$~keV and a total luminosity (being observed face-on at 
$r > r_1 = 1.7 \cdot 10^{11}$\,cm) of $L_{BB} \sim 3 \cdot 10^{37}$\,erg/s.
The soft excess may be also fitted with a multicolour funnel (MCF) model
(\cite[Fabrika \etal\ 2006]{Fab_etal06}).
The soft excess is observed in the spectra of ULXs with the soft component
temperature of $T \sim 0.1$~keV (\cite[Stobbart, Roberts, \& Wilms 2006]{Stobbart06}), 
which is similar to that found 
in SS\,433. If the ULXs or some of them are nearly face-on versions of 
SS\,433, one may adjust their soft X-ray components to the outer funnel 
walls radiation (\cite[Poutanen \etal\ 2007]{Poutan07}).

We conclude that the additional reflected spectrum is an indication of 
the funnel radiation. The illuminating radiation spectrum is flat in the range 
of 7\,--\,12~keV, as it is expected in supercritical accretion disks (\cite[Poutanen \etal\ 2007]{Poutan07}). 
Comptonization may extend and flatten the spectrum to higher energies.
With multiple scatterings in the funnel the hard radiation
may survive absorption. The observed reflected luminosity in this 7\,--\,12~keV
spectral range is $L_{refl} \sim 10^{36}$\,erg/s. We find that the illuminating 
luminosity has to be no less than  $\sim 10^{39}$\,erg/s (\cite[Medvedev \& Fabrika 2010]{Medv2010}).

In the range 2.0\,--\,7.0~keV the additional reflected spectrum is curved 
(Fig.\,\ref{fig3}). Therefore we expect an existence of an absorbing 
(reflecting) medium inside the funnel, which produces such a curved
spectrum of radiation, illuminating the outer reflected surface. Presumably
this medium is the deep funnel regions.
The recent data show (\cite[Stobbart, Roberts, \& Wilms 2006]{Stobbart06}; 
\cite[Berghea \etal\ 2008]{Berghea08}) that ULXs posses curvature and rather flat 
X-ray spectra, which are difficult to interpret with a single-component or
any other simple model. We note that the main properties of the ULXs X-ray spectra 
are similar to those restored additional spectral components in SS\,433, the ULXs may 
be "face-on" versions of SS\,433.
 
The softer (2\,--\,7~keV) part of the illuminating spectrum (Fig.\,\ref{fig3}) 
carries a trace of absorption. This is assumed to be due to the multiple scatterings in 
the funnel. It is important to note that an existence of He- and H-like absorption edges 
at about zero velocity is a mandatory property of the funnel spectrum to be able to
produce the observed jet velocity due to the line-locking mechanism.
The jet velocity value, $v_j \approx 0.26 c$, and its unique stability,
where the velocity does not depend on the activity state, indicate that the jet 
acceleration must be controlled by the line-locking mechanism 
(\cite[Shapiro, Milgrom, \& Rees 1986]{shapiro86}; \cite[Fabrika 2004]{Fab04}).

\end{document}